Realization of macroscopic ratchet effect based on nonperiodic and uneven potentials.

V. Rollano,[1] A. Gomez,[2] A. Muñoz-Noval,[3] M. Velez,[4,5] M. C. de Ory,[1] M. Menghini,[1] E. M. Gonzalez,[1,3] and J L Vicent[1,3.a)]

[1] IMDEA-Nanociencia, Cantoblanco, E-28049 Madrid, Spain

[2] Centro de Astrobiología (CSIC-INTA), Torrejón de Ardoz, E-28850 Madrid, Spain

[3] Departamento Fisica de Materiales, Universidad Complutense, E-28040 Madrid, Spain

[4] Departamento de Física, Universidad de Oviedo, E-33007 Oviedo, Spain.

[5] CINN (Universidad de Oviedo-CSIC), E-33940 El Entrego, Spain.

[a)] E-mail: jlvicent@ucm.es

Abstract

Ratchet devices allow turning an ac input signal into a dc output signal. A ratchet device is set by moving particles driven by zero averages forces on asymmetric potentials. Hybrid nanostructures combining artificially fabricated spin ice nanomagnet arrays with superconducting films have been identified as a good choice to develop ratchet nanodevices. In the current device, the asymmetric potentials are provided by charged Néel walls located in the vertices of spin ice magnetic honeycomb array, whereas the role of moving particles is played by superconducting vortices. We have experimentally obtained ratchet effect for different spin ice I configurations and for vortex lattice moving parallel or perpendicular to magnetic easy axes. Remarkably, the ratchet magnitudes are similar in all the experimental runs; i. e. different spin ice I configurations and in both relevant directions of the vortex lattice motion. We have simulated the interplay between vortex motion directions and a single asymmetric potential. It turns out vortices interact with uneven asymmetric potentials, since they move with trajectories crossing charged Néel walls with different orientations. Moreover, we have found out the asymmetric pair potentials which generate the local ratchet effect. In this rocking ratchet the particles (vortices) on the move are interacting each other (vortex lattice); therefore, the ratchet local effect turns into a global macroscopic effect. In summary, this ratchet device benefits from interacting particles moving in robust and topological protected type I spin ice landscapes.

INTRODUCTION

Ratchet effect is at the core of notable and unique phenomena, for example biological molecular machines (1) and Brownian motors (2). This effect happens when an ensemble of particles, driven by zero average forces is moving on asymmetric potentials. Ratchet mechanism yields a directional net motion of the particles despite the applied driven force is not. So, a rectifier effect happens: an ac input signal yields a dc output signal. Different types of ratchets can be studied, for example flashing or tilting ratchets. We focus on tilting, deterministic, rocking ratchets. In this situation, ratchet potentials are not time-dependent and the zero-average driving force is periodic. In the usual case of periodically arranged asymmetric potentials the effect of each potential adds (3). Magnetic topological defects have been used as an appropriate tool in ratchet mechanisms (4-7). The combination of superconducting films on top of array of nanomagnets has been shown very useful to capture the rich and complex physics of the ratchet effect (8).

In the present work, the asymmetric potentials are provided by a honeycomb array of connected nanomagnet bars. These arrays develop a set of frustrated nanomagnets, which are governed by spin ice rules, see the reviews (9, 10). In these spin ice structures many different topics have been addressed, as for example magnetic monopoles (11, 12), magnetic chiral order (13) and ratchet systems (14-17). We have to underline that, due to the interaction between the magnetic dipoles at the end of the bars, a distribution of positive (+1) and negative (-1) spin ice charges emerges (18). In thin nanostructures with holes, as are honeycomb arrays shaped by connected stripes, the micromagnetic characterization shows fractional magnetic vortices and composite magnetic walls (17, 19-20). In particular, in our case of connected nanomagnets bars, two charged Néel walls are developed at each vertex of the array and supply the asymmetric potentials. In summary, in this hybrid system (superconducting film on top of the honeycomb array), when an ac current is injected, the superconducting vortices move on asymmetric potentials caused by the Néel wall pairs. The outcome is a net flow of vortices; therefore, a dc voltage is measured (17).

This ratchet effect has been studied in two situations: i) at remanence, after an in-plane saturating magnetic field is applied along one of the three anisotropy easy axes, i.e, an axis parallel to one of the directions of the nano-bars. When the saturating field is decreased to zero, a periodic and ordered distribution of positive and negative charges shows at remanence. The sample is in spin ice II configuration. ii) at remanence, when a saturating magnetic field is applied along one of the three anisotropy hard axes (perpendicular to one of the easy axes), then the positive and negative magnetic charges distribution is disordered at remanence. The sample is in spin ice I configuration.

In the first situation, spin ice II configuration, the periodic and ordered asymmetric potentials yield a ratchet effect. Remarkably, in the disordered case, spin ice I configuration, a ratchet effect is also observed. This noticeable ratchet effect is supported by the fact that a non-zero remanent magnetization exists in the disordered state (Ice I). This happens since, when the saturating field applied along the hard axis is decreased to zero, this magnetic field seems to imprint a preferential direction in the Ice I configuration. The final result is not a fully disordered set of asymmetric potentials since the system remembers the preferential direction defined by the saturation field (18).

In the present paper, we will go a step further, since we experimentally explore different type I configurations and different vortex motion trajectories and we analyze the local distribution of asymmetric potentials. We show that in this hybrid system ratchet effect occurs whichever is the distribution of asymmetric potentials; i. e. any configurations of charges in this type I spin ice yields a ratchet effect. Moreover, we have modeled the local distribution of asymmetric potentials, to figure out which pairs of asymmetric potentials are relevant to promote the ratchet effect and which of the pairs compensate each other, being irrelevant for the ratchet effect.

EXPERIMENTAL METHODS

The hybrid nanostructured system comprises a honeycomb array of magnetic Co nanobars (Fig. 1 (a)) embedded in a Nb superconducting film. This hybrid sample is made by a combination of electron beam lithography, sputtering and etching techniques, on Si substrates (17).The dimensions of the hexagonal unit cell are side length 300 nm, width 150 nm and thickness 20 nm. On top of the array, a 100 nm thick Nb film was deposited using magnetron sputtering. The hybrid sample is patterned to a cross-shaped bridge (Fig. 1 (b)). Therefore, the current can be applied parallel or perpendicular to one of the sample magnetic easy axes, which are parallel to the Co bars. The sample is placed in a He cryostat with a superconducting solenoid and temperature controller with temperature stability better than 1 mK. The sample is fixed in a computer controlled rotatable sample holder which allows applying the magnetic fields in-plane or out-of-plane. The usual four-probe dc technique is used to record the data. We have to point out that we are dealing with adiabatic rocking ratchet, therefore the effect is independent of the frequency (21). When the injected (ac) alternating (1 kHz frequency) input current is applied along one of the hard (easy) axes, the vortices move perpendicularly along one of the easy (hard) axes (8). The output dc voltages are recorded with a commercial nanovoltmeter.

The experiment protocol is summary as follows: we set the ac current density which yields an ac Lorentz force on the vortices that is given by $\mathbf{F_L} = \mathbf{J} \times \mathbf{n} \, \Phi_0$, where the current density is $J = J_{ac} \sin \omega t$, $\mathbf{n}$ is a unitary vector parallel to the applied magnetic field, and $\Phi_0$ is the fluxoid. The vortex lattice flows in an asymmetric potentials landscape placed in the vertices of the honeycomb array. From the Josephson expression for the electric field $\mathbf{E} = \mathbf{B} \times \mathbf{v}$, being $\mathbf{B}$ the applied magnetic field and $\mathbf{v}$ the vortex-lattice velocity, we measure a dc voltage drop $V_{dc}$ that is given by $V_{dc} = v B d$, where $d$ is the distance between contacts and $B$ the applied magnetic field. For an ac driving Lorentz force we find that the output is a nonzero dc voltage $V_{dc}$. This means that a net vortex lattice flow arises from the ac driving force. Albeit the time averaged force on the vortices is zero ( $<F_L>$ = 0) a nonzero dc voltage drop is measured ( $V_{dc} \neq 0$ ). In summary, it turns out a ratchet effect is observed.

Magnetic Force Microscopy (MFM) contrast was used as a proxy for the potential asymmetric landscape of the nanomagnets. Contrast profiles of the asymmetric potentials have been extracted from experimental and simulated MFM data. The experimental distribution of asymmetric Néel walls was obtained by using a NANOTECH Atomic Force Microscope system with magnetic NANOSENSORS (22). The micromagnetic simulations were carried out with the finite difference code MuMax3 (23) . Typical Co parameters have been used: $M_s$ = 1.4 x $10^6$ A $m^{-1}$, A = 3 x $10^{-11}$ J $m^{-1}$ and K = 0 J $m^{-3}$, being $M_s$ the saturation magnetization, A the exchange constant, and K the in-plane anisotropy. Polycrystalline cobalt magnetocrystalline anisotropy was neglected due to its weak strength in comparison to shape anisotropy.

RESULTS AND DISCUSSION

We have studied two samples (A and B) with exactly the same honeycomb array (Methods). Sample A is measured as-grown ($T_{c0}$ = 8.91 K); i. e. the only applied magnetic fields are the weak fields applied perpendicularly to the sample plane, which are needed to generate the superconducting vortices. Morgan et al. (24) have studied the magnetic state of as-fabricated artificial magnetic square ice and they found that the expected ground state is very difficult to reach, since the energy barriers to thermal equilibrium are very large. These authors identify the state as the frozen-in residue of true thermodynamics that occurred during the fabrication of the sample. We can expect in our case something similar, i. e. to work out in detail this as-grown magnetic state could be rather speculative.

Figure 2 (a) depicts a sketch of the experimental layout, which is the same that was reported in Ref. 17 ; see Methods and the Supplemental Information for more details. Magnetic Force Microscopy (MFM) allows obtaining the experimental distribution of +1 and -1 magnetic charges in sample A and B. Figure 2(b) shows an experimental MFM image of the as-grown sample (sample A). The distribution of +1 and -1 magnetic charges does not show any specific pattern; the sample is in spin ice I configuration. In sample B ($T_{c0}$ = 8.4 K) a different type I configuration is realized following a magnetic protocol, that we can call remanent state selection protocol. Firstly, a 7 T saturating magnetic field is applied along one of the easy axes. Once the field is removed, the sample is in spin ice II configuration at remanence, with an ordered distribution of the magnetic charges. Then, a 7 T saturating field is applied parallel to a hard axis which is perpendicular to the former easy axis. After removing the field, the remanence state of the spin ice belongs in type I configuration. In the remanent state selection protocol, the spin ice retains some memory of the applied magnetic field. Figure 2(c) shows an experimental magnetic force microscope (MFM) image of the sample B at remanence. More details can be found in Supplemental Information.

In both samples (A, B) ratchet effect has been studied taking into account two vortex motion directions: Superconducting vortices on the move parallel to one of the magnetic easy axes (array nanobars) and parallel to one of the magnetic hard axes (perpendicular to one of the easy axes); more details in Supplemental Information.

In sample A, spin ice I has been achieved without imposing a chosen magnetic memory direction. We can study rocking ratchet effect behavior in as-grown sample; i. e. magnitude and temperature dependence and whether or not this behavior deviates from the reported and expected behavior in this type of ratchet. Conversely, in sample B a privilege direction has been set by means of a remanent state selection protocol. Hence, we can study whether or not this ratchet effect can discriminate the preferent magnetic direction imprinted in the device.

Figure 3 shows the temperature dependence of ratchet effect measured in sample A (as-grown) with a perpendicularly applied field $H_1$ = 40 Oe, which corresponds to vortex density of one vortex per triangular unit cell (17). Figure 3 (a) depicts ratchet effect when the vortex motion is set parallel to one of the easy-axes. Figure 3 (b) depicts ratchet effect when the vortex motion is set parallel to one of the hard-axes. The ratchet effect shows the usual and expected behavior with temperature; i. e. decreasing temperature induces an enhancement of ratchet amplitude (25). We have experimentally obtained a ratchet effect in a system with the vortices moving in a disorder and unknown distribution of asymmetric magnetic potentials. In the as-grown configuration there is not a chosen preferred magnetic direction, since no magnetization protocol has been applied at all. So, this ratchet effect emerges from the combination of local asymmetries distributed along the sample.

Otherwise, in sample B a remanent state magnetic selection protocol has been applied. Two noticeable directions can be explored for vortex lattice motion. One of them when vortex lattice moves parallel to one of the easy axes (Co nanobars), and the second one when the vortices are on the move perpendicular to the easy axis (see Supplemental Information). Figure 4 shows the experimental ratchet results in both situations. This peculiar ratchet effect seems to happen from specific distributions of asymmetric potentials, whatever the global state of spin ice I. This outcome can be explored in two steps, firstly we study different vortex trajectories through a single potential, to discriminate which of them do produce a ratchet signal and which do not produce a ratchet effect. Then we have to find out the combination of pairs of asymmetric potentials which are relevant to the ratchet effect. In summary, to understand the experimental results we have to investigate the vortex trajectories and the selective pairs of local asymmetric potentials that produce the local ratchet effect and then we have to clarified why this local ratchet effect does not vanish when we probe the whole sample.

We have to take into account the main features of these asymmetric potential wells which comprise two charged Néel walls. In the honeycomb spin-ice nanomagnets, there are -1/2 edge magnetic half vortex with positive magnetic ice charge and -1/2 edge magnetic half vortex with negative magnetic ice charge (19, 20). Figure 5 panels (b) and (c) show simulated MFM profiles of the Néel walls. These profiles have been obtained following the methodology describes in Experimental Methods, which allow monitoring asymmetry changes when the vortex trajectories deviate from a perfect symmetrical profile; i. e. panel (a) (0º). Panels (b) and (c) correspond to trajectories at 2º and 10º with respect to the symmetric situation as indicated in panel (d), in both cases the vortex trajectories cross over asymmetric potentials . Therefore, the asymmetric potential condition can be accomplished very easily for different vortex trajectories. It is crucial the orientation of vortex trajectory with the two charged Néel walls, which shape the asymmetric potentials.

Next, the combinations of asymmetric potentials are studied. In the ordered magnetic state (Ice II configuration), the positive and negative ice charges are alternating (17) (see Supplemental Information). The superconducting vortices are moving in a locally asymmetric potential landscape and each (+1 , -1) potential pair pushes the vortex lattice in the same direction and a net flow happens, since both potentials add each other. In principle, only alternating (+1, -1) potentials add, while (-1, -1) or (+1, +1) charges in a row cancel out and do not contribute to the ratchet effect. Hence, in the Ice I configuration the disordered potential landscape precludes this straightforward analysis. To tackle this issue, we have simulated spin ice I (see Supplemental Information) ; from this simulation we have obtained the potential profiles and we have compared them to the experimental ones obtained from the MFM scans. In this way, we have a double check to discard the pair potentials which does not contribute and we will be able to find out the local potential pairs which sum up to ratchet effect. For this analysis we have chosen a simulated unit cell at random (see Fig. 6(a)). Figures 6(b)-(d) show the potential profiles across this unit cell from the simulated array, where two out of three trajectories are crossing the same polarity spin ice charges in a row. A similar analysis is also done in pairs of spin ice charges found in the experimental spin ice I configuration, see Figs. 6(e)-(g). Remarkably, we observe that uneven potentials with the same charge polarity can trigger the ratchet effect, see Figs. 6(c) and 6(f), since the asymmetric potentials do not cancel, they add each other for certain vortex trajectories. The opposite happens in trajectory 3 where both asymmetric potentials cancel each other (Figs. 6(d) and 6(g)).

This shows that specific combination of potential well profiles is crucial for the emergence of the ratchet effect at large scales. Many asymmetric potential pair candidates are present and they can generate and contribute to this robust ratchet effect; both with opposite spin ice charges and with the same spin ice charges. Furthermore, since superconductivity is a cooperative long-range order effect, we are dealing with a ratchet of interacting particles, which are sculpted in a lattice, the vortex lattice. Therefore, the local ratchet effect becomes a macroscopic ratchet effect. The vortex lattice interplays with the asymmetric potentials and the interacting nature of superconducting vortices are crucial to obtain a macroscopic net flow of particles (vortices) and a rectifier effect from the disordered distribution of asymmetric potentials. There are non-negligible potential asymmetries for superconducting vortex motion for many different magnetic vertex configurations. Our experimental results suggest that collective interactions within the superconducting vortex lattice enhance the effect of the local and uneven asymmetries. A steady ratchet effect is generated in the disordered magnetic potentials in spin ice I configurations with different remanent magnetization.

CONCLUSIONS

In summary, our results show that a rocking ratchet effect is the result of the asymmetries provided by the frustrated nanomagnets in the honeycomb array, even when a magnetic preferential direction is not identified. Through MFM experiments and simulations, we have proved that, a rich set of local pairs of asymmetric potentials exists, which contribute to the ratchet effect even with the same charge polarities. Therefore, a rectifier device can be achieved of interacting particles (vortices) regardless the distribution of the magnetic asymmetric potentials and the strengths of their potential wells. Finally, this macroscopic ratchet effect does not show any specific dependence with spin ice I configurations neither the vortex lattice motion direction; i. e. parallel or perpendicular to magnetic easy axes (bars in the honeycomb array).

ACKNOWLEDGMENTS

This work was supported by Spanish MICINN grants FIS2016-76058 (AEI/FEDER, UE), EU COST-CA16218. IMDEA Nanociencia acknowledges support from the 'Severo Ochoa' Programme for Centres of Excellence in R&D (MICINN, Grant SEV-2016-0686). MCO and AG acknowledges financial support from Spanish MICINN Grant ESP2017-86582-C4-1-R and IJCI-2017-33991; AMN acknowledges financial support from Spanish CAM Grant 2018-T1/IND-10360. MV acknowledges financial support from Spanish MICINN Grant PID2019-104604RB/AEI/10.13039/50110001103.

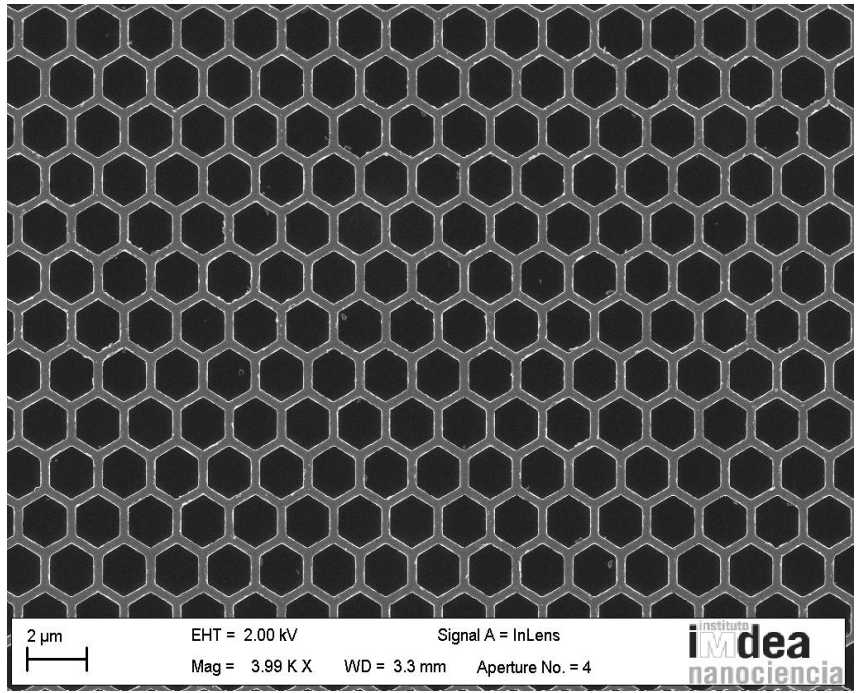

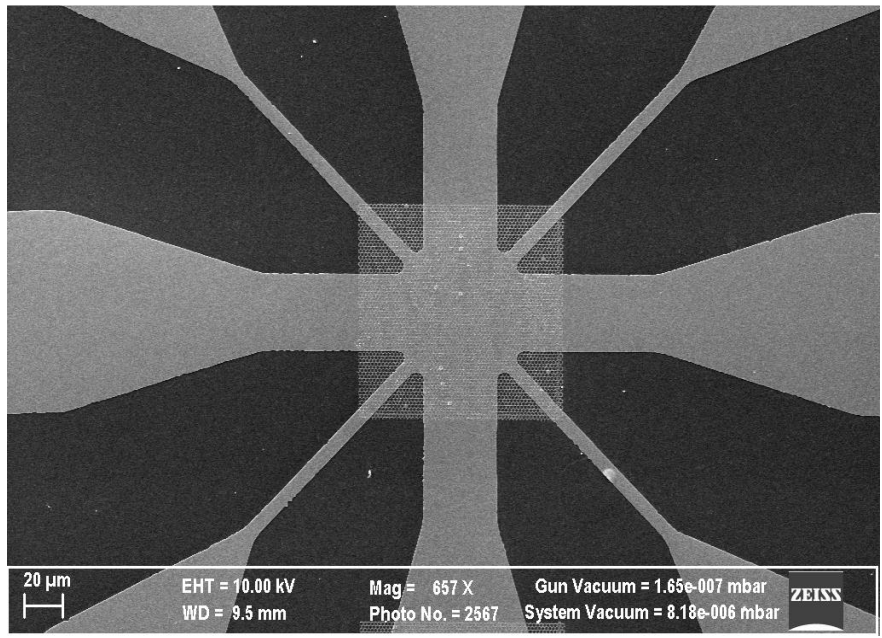

**Figure 1**. (a) Scanning microscope image of the Co honeycomb array. (b) Scanning microscope image of cross-shaped bridge (Nb film on top of Co honeycomb bridge).

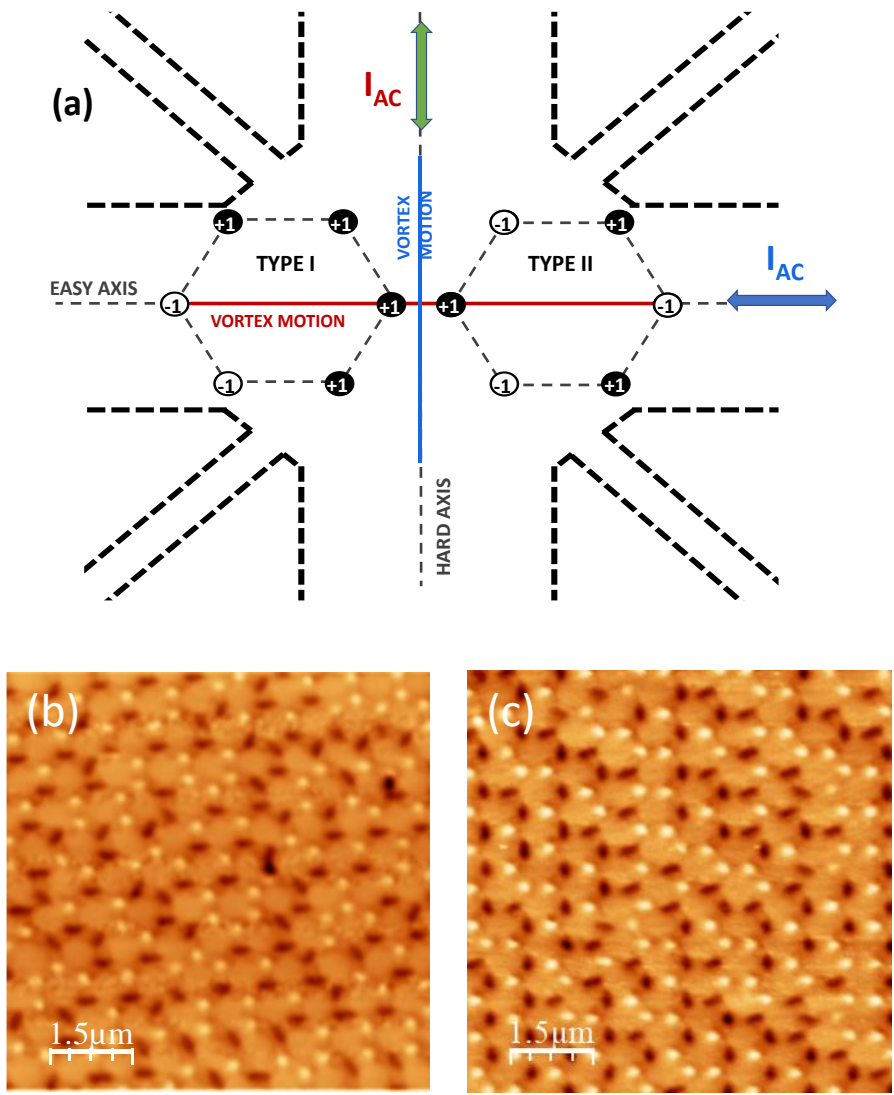

**Figure 2.** (a) Experimental configuration sketch. Including cross-shaped bridge (see Methods) and the typical spin ice configurations in the honeycomb array (in this work only type I is studied). Colors red and blue identify the two directions of motion for superconducting vortices, which are perpendicular to the AC driving current (double-headed arrows). Black and white dots represent magnetic charges from the spin ice providing the asymmetric landscape that yields ratchet effect. (b) Experimental MFM image of as-grown sample (sample A) and (c) Experimental MFM image of remanent state magnetic selection protocol sample (sample B); see text .

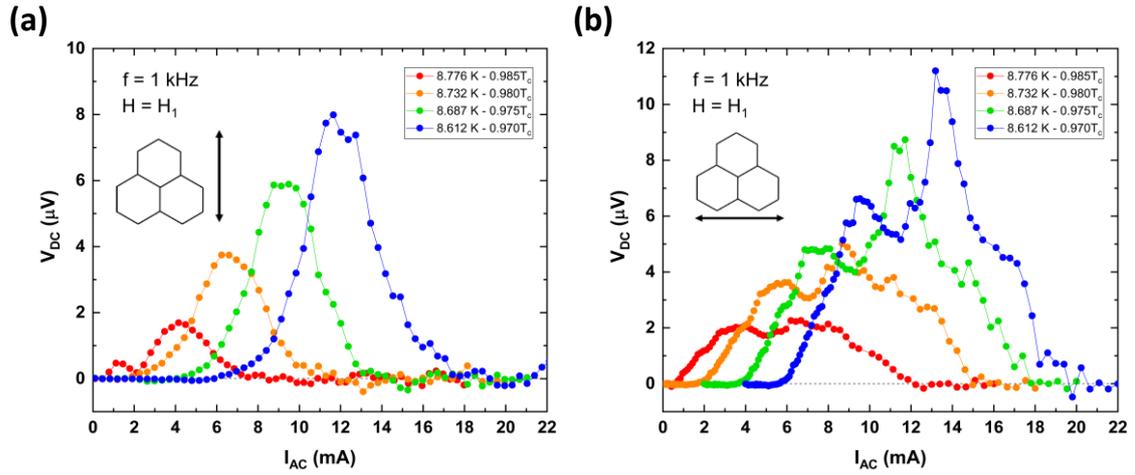

**Figure 3**. Temperature dependence of ratchet effect in as-grown (sample A) Nb film on top of honeycomb spin ice of Co connected nanomagnets (Temperatures are shown in the inset). Y-axis output dc voltages. X-axis input ac current (1 kHz). The applied magnetic field perpendicular to the sample plane is $H_1$ = 40 Oe (filling factor: one vortex per array unit cell) and critical temperature $T_{c0}$ = 8.91 K. (a) Vortex motion parallel to easy-axis (inset shows the honeycomb array and direction of the vortex motion). (b) Vortex motion parallel to hard-axis (inset shows the honeycomb array and direction of the vortex motion).

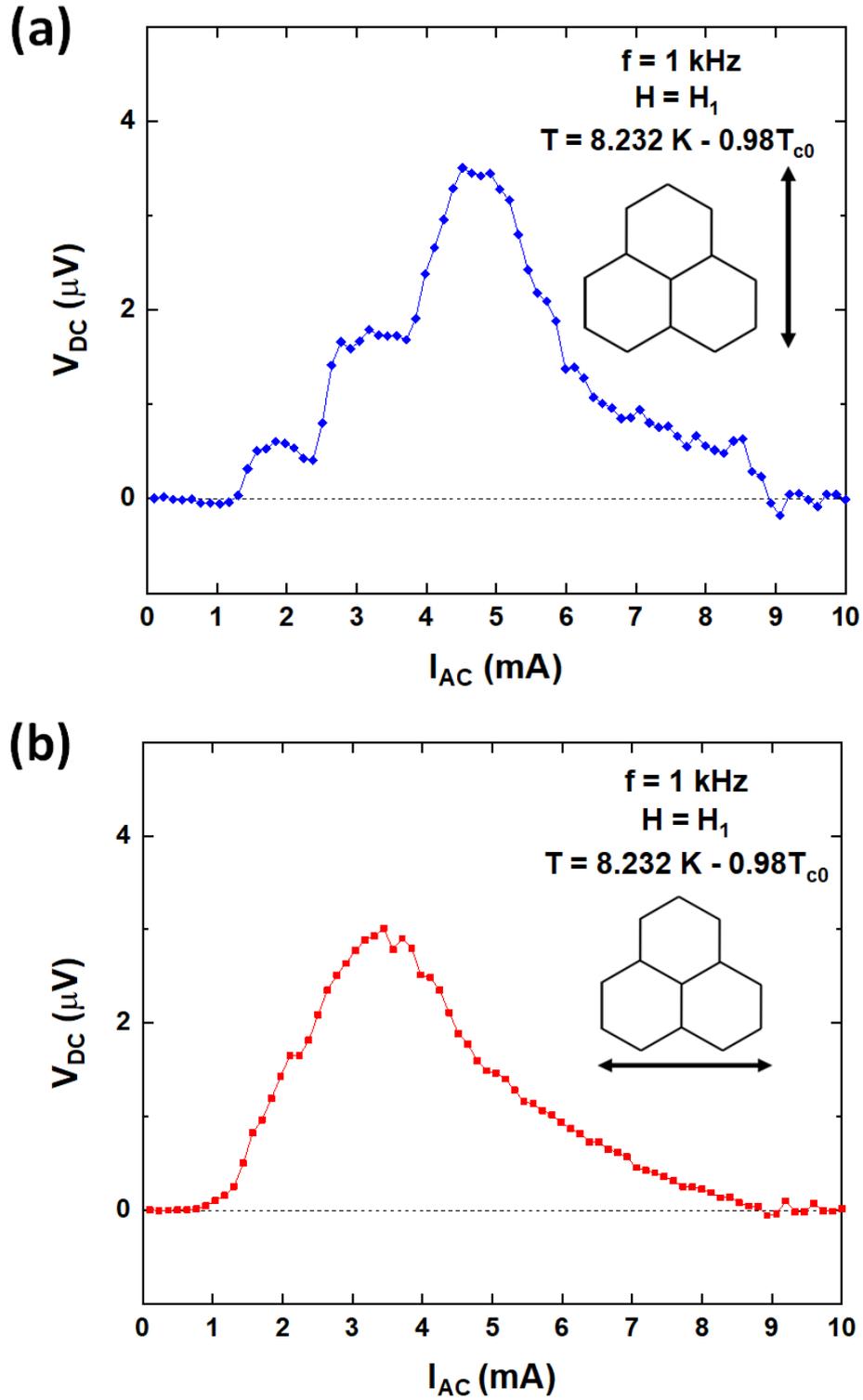

**Figure 4**. Ratchet effect at $T/T_{c0}$ = 0.98 of sample B, applied magnetic field out of sample plane $H_1$ = 40 Oe (filling factor: one vortex per array unit cell). (a) Vortices moving parallel to easy axis. (b) Vortices moving parallel to hard axis. Insets show the direction of vortex motion.

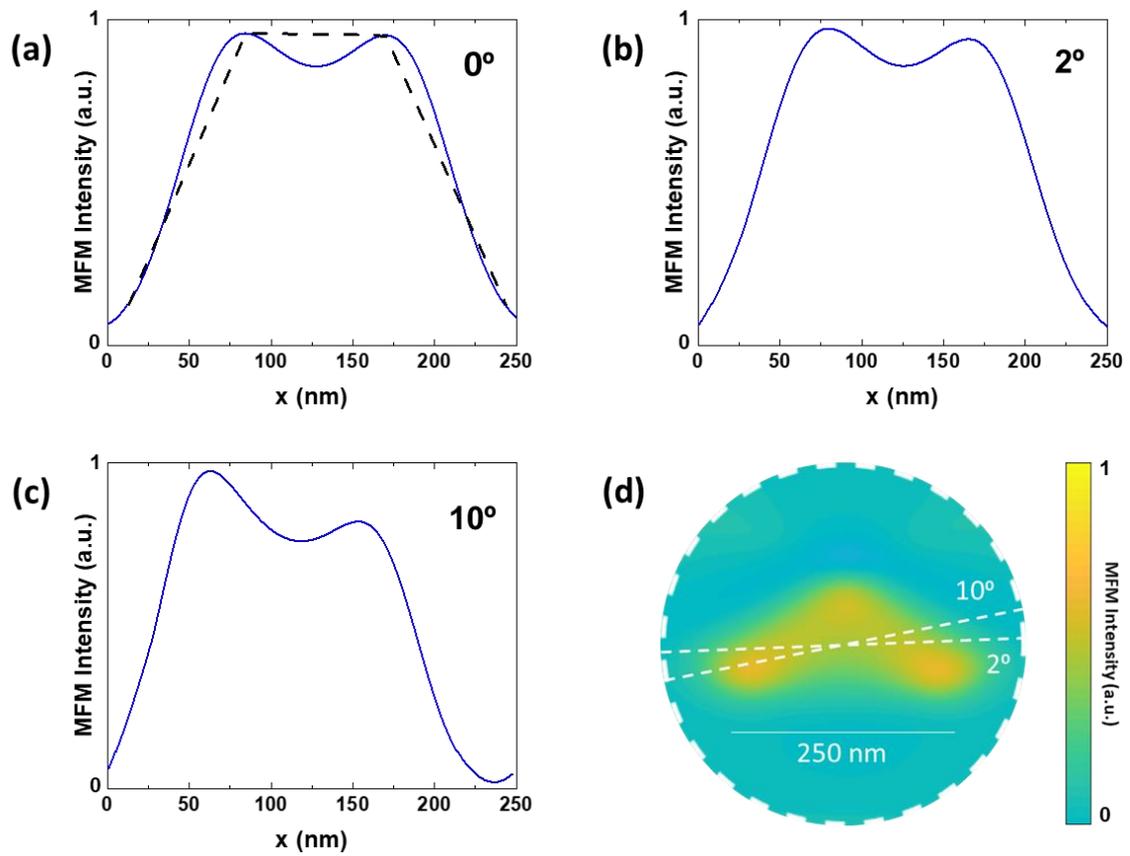

**Figure 5**. Simulated profiles of a single -1 magnetic charge potential well for different vortex trajectories, to monitor asymmetry changes as the vortex movement deviates from a perfectly symmetrical profile (0 deg) (a); (b) corresponding to 2 deg.; (c) corresponding to 10 deg. ; (d) sketch of the asymmetric potential with two vortex trajectories depicted.

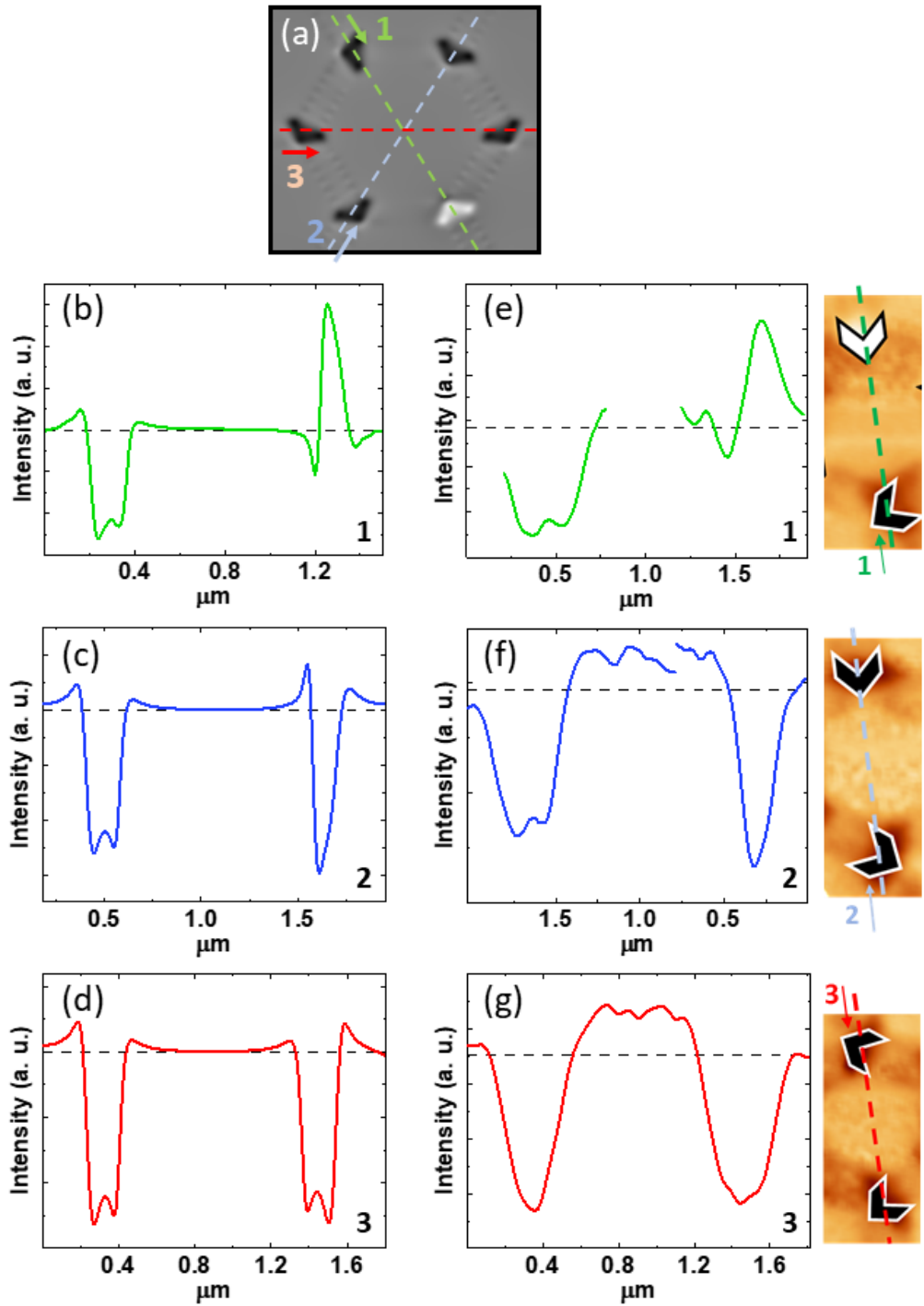

Figure 6. (a) Hexagonal unit cell extracted from simulation of the spin ice array in Ice I configuration. The black and white nanotriangles correspond to +1 and -1 spin ice charges, respectively. The 1, 2, 3 dashed lines correspond to possible vortex trajectories. (b)-(d) simulated profiles along 1, 2 and 3 trajectories respectively. (e)-(g) Experimental profiles from experimental MFM images. They are taken from the MFM images placed to the right of the respective panels. The (e), (f) and (g) panels correspond to the same (Fig. 6(a)) trajectories 1, 2, 3 as the ones in the simulated profiles Figs. 6(b)-(d).

SUPPLEMENTAL INFORMATION

Realization of macroscopic ratchet effect based on nonperiodic and uneven potentials.

V. Rollano,[1] A. Gomez,[2] A. Muñoz-Noval,[1,3] M. Velez,[4,5] M. C. de Ory,[1] M. Menghini,[1] E. M. Gonzalez,[1,3] and J. L. Vicent[1,3,a)]

[1] IMDEA-Nanociencia, Cantoblanco, E-28049 Madrid, Spain

[2] Centro de Astrobiología (CSIC-INTA), Torrejón de Ardoz, E-28850 Madrid, Spain

[3] Departamento Física de Materiales, Universidad Complutense, E-28040 Madrid, Spain

[4] Departamento de Física, Universidad de Oviedo, E-33007 Oviedo, Spain.

[5] CINN (Universidad de Oviedo-CSIC), E-33940 El Entrego, Spain.

a) E-mail: jlvicent@ucm.es

The magnetic configuration of the Co honeycomb lattice is described in terms of two kinds of −1/2 magnetic half vortices, either associated with a +1 magnetic charge (black half vortex) or with a −1 magnetic charge (white half vortex); each vertex contains two charged Néel walls.

The orientation of each half magnetic vortex in each vertex of the honeycomb array can be extracted taking into account the half magnetic vortex asymmetries and following the pseudo spin ice rules. That is: Two in - one out (+1 magnetic charge in the vertex) or two out - one in (-1 magnetic charge in the vertex) (1). In this way, we can have a picture of the whole sample with the +1 y -1 magnetic charge distribution.

Figures S1(a) - (c) depict: (a) MFM experimental image, (b) simulated MFM contrast and (c) simulated micromagnetic configuration. Comparison between (a-c) allows determining the position of ice charges +1 (black) and -1 (white) as sketched in Fig. 1 (d).

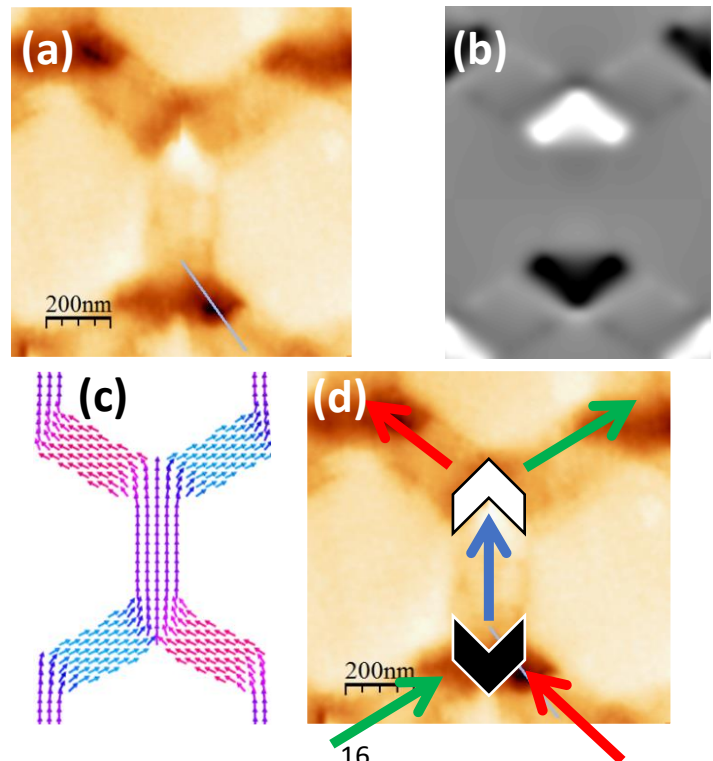

**Figure S1**. (a) Experimental image of a pair (-1, +1) of spin-ice charges (corresponding to magnetic half vortices) in a single bar of the array, (b) simulated MFM contrast and (c) simulated micromagnetic configuration, showing the two Néel walls in each vertex. Comparison between (a) and (c) allows determining the position of magnetic half vortices and the local magnetization orientation at each intersection as sketched in (d).

Figure S2 shows typical experimental MFM image of the honeycomb array in spin-ice I configuration. There is not any specific magnetic charge pattern in the picture.

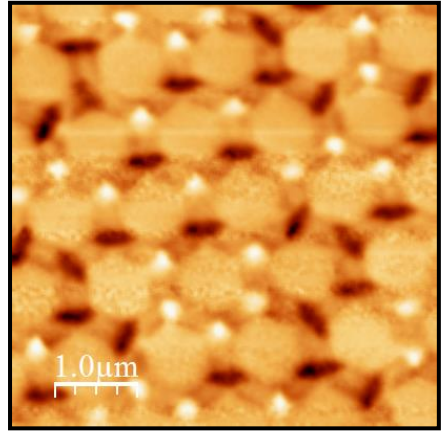

**Figure S2** Experimental Magnetic Force Microscopy (MFM) image of the honeycomb array in spin-ice I configuration.

Following the pseudo spin ice rules, the magnetization directions can be known in the honeycomb bars. Figure 3 depicts the outcome.

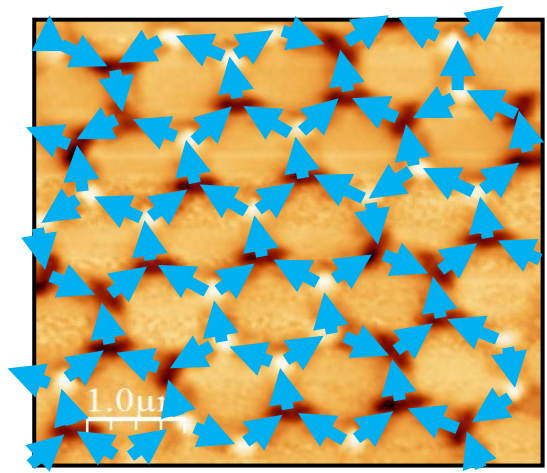

**Figure S3.** Magnetic dipole directions after the pseudo spin ice rules are applied to Figure S2 (MFM experimental image).

Worth to note that the V-shaped pairs of charged Néel domain walls of these two half vortices point in different directions along the honeycomb array in this spin ice I configuration; see

Figure S4. In this Figure the black V-shaped corresponds to +1 magnetic charge and the white V-shaped corresponds to -1 magnetic charge.

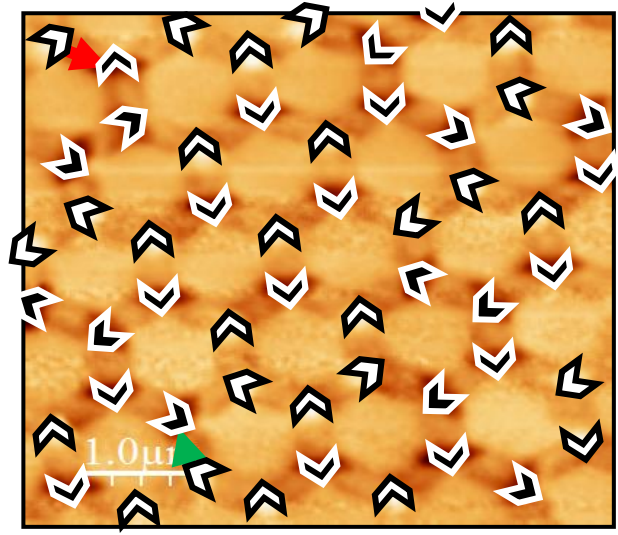

**Figure S4**. Distribution and orientation of the +1 (black) and – 1 (white) magnetic charge.

Therefore, a vortex on the move can cross different orientations of the two charged Néel walls, following the same trajectory.

In addition, this analysis allows a rough estimation of the direction of the remanent magnetization in both samples. Figure S5 shows a sketch.

<u>Sample A:</u>

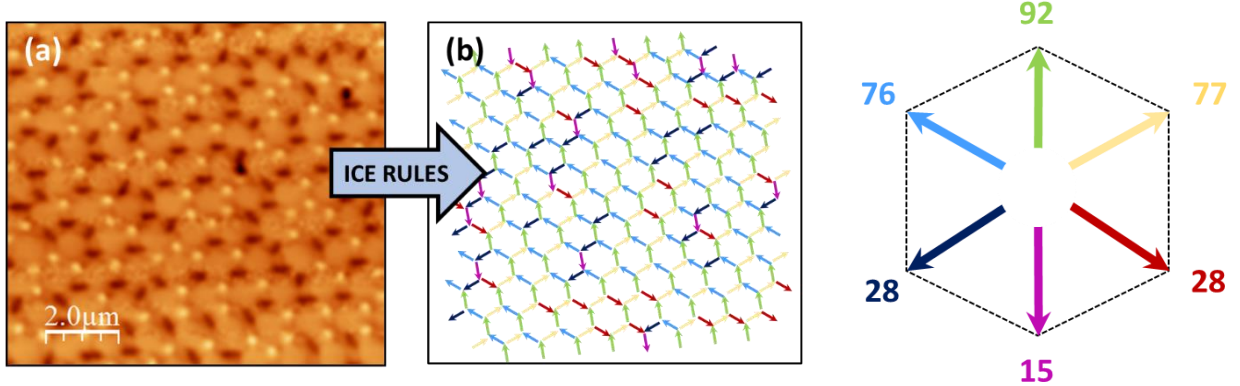

Sample B

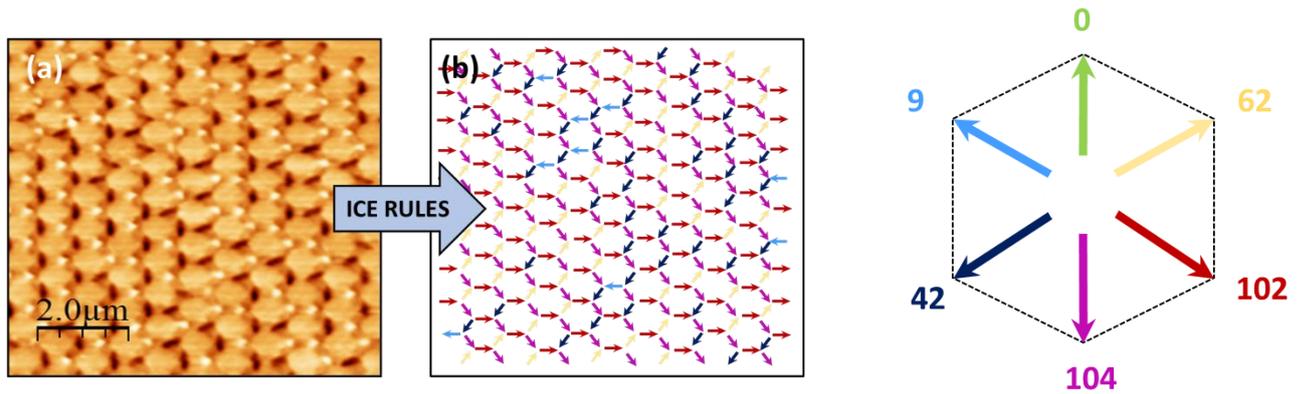

**Figure S5.** (a) Sample A. Magnetization distribution obtained from MFM experimental image following the ice rules. (b) Sample B. Magnetization distribution obtained from MFM experimental image following the ice rules.

Interestingly, in sample A the resultant total magnetization is pretty close to an easy axis direction, while in sample B the total magnetization is deviated from any symmetrical axes and with strength much higher than in sample A.

Figure S6 shows the difference between spin ice I (disordered) and spin ice II (ordered) configurations.

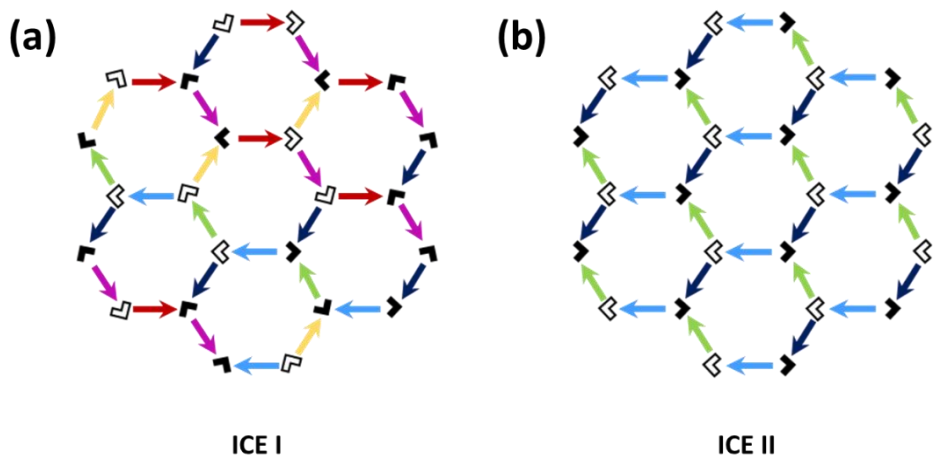

**Figure S6.** Examples of spin ice configurations. (a) Type I spin ice. (b) Type II spin ice.

Finally, in Figure S7 a sketch of the experiment geometry is depicted.

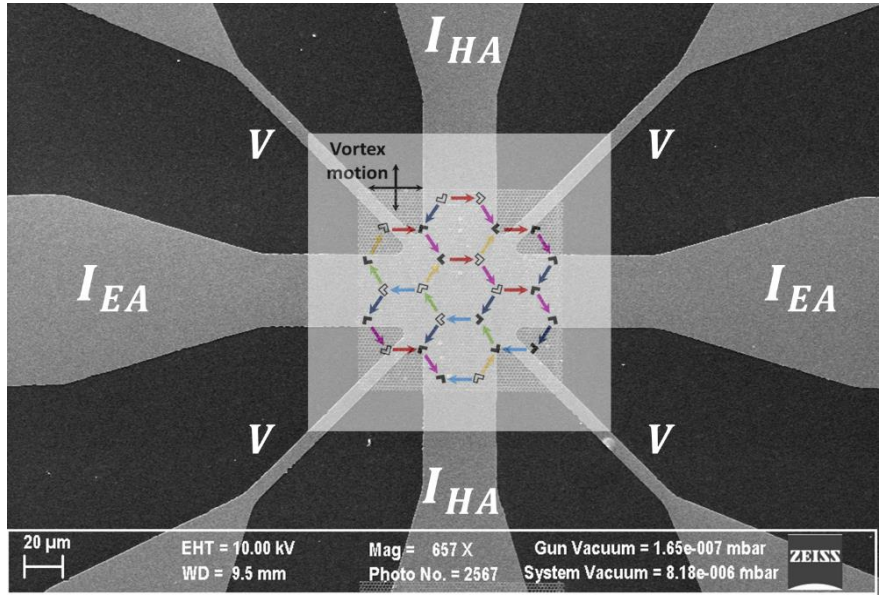

**Figure S7.** SEM image of the device. A drawing of the honeycomb array has been inserted. $I_{EA}$ and $I_{HA}$ indicate the current directions parallel to easy and hard axes respectively.

**References**

1	Tanaka, M., Saitoh, E., Miyajima, H., Yamaoka, T. & Iye Y. Magnetic interactions in a ferromagnetic honeycomb nanoscale network. *Phys. Rev. B* **73**, 052411 (2006)

2121